\documentclass[journal]{IEEEtai}

\usepackage[colorlinks,urlcolor=blue,linkcolor=blue,citecolor=blue]{hyperref}

\usepackage{color,array}
\usepackage{graphicx}
\usepackage{amsmath}
\usepackage{amssymb}
\usepackage{xcolor}
\usepackage{url}
\usepackage{booktabs}   
\usepackage{multirow} 
\usepackage{comment}

\begin{document}

\title{Decision-Theoretic Safety Assessment of Persona-Driven Multi-Agent Systems 
in O-RAN}

\author{Zeinab~Nezami,
        Syed~Ali~Raza~Zaidi,
        Maryam~Hafeez,
        Louis~Powell,
        Vara~Prasad~Talari,
        and Mallik~Tatipamula%
\thanks{Z. Nezami is AI Data Engineering Lead at the GSMA.}%
\thanks{S. A. R. Zaidi and M. Hafeez are with the School of Electronic and Electrical Engineering, University of Leeds.}%
\thanks{L. Powell is Director of AI Initiatives at the GSMA.}%
\thanks{V. P. Talari is Principal Technologist at AWS.}%
\thanks{M. Tatipamula is Chief Technology Officer at Ericsson.}%
}

\maketitle

\begin{abstract}
Autonomous network management in Open Radio Access Networks requires intelligent decision-making across conflicting objectives, yet existing LLM-based multi-agent systems employ homogeneous strategies and lack systematic pre-deployment validation. We introduce a persona-driven multi-agent framework where configurable behavioral personas—structured specifications encoding optimization priorities, risk tolerance, and decision-making style—influence five specialized agents (planning, coordination, resource allocation, code generation, analysis). To enable rigorous validation, we develop a three-dimensional evaluation framework grounded in decision theory, measuring normative compliance (optimality adherence), prescriptive alignment (behavioral guideline consistency), and behavioral dynamics (emergent system properties).

We evaluate 486 persona configurations across two O-RAN optimization challenges (energy-efficient resource allocation and network load balancing). Results demonstrate that persona-agent alignment significantly impacts both individual performance (±14.3\% range) and emergent multi-agent coordination, with retrieval architecture (GraphRAG vs. RAG) fundamentally constraining customization effectiveness. Single-agent persona modifications propagate system-wide through cascading effects, with certain combinations exhibiting detectable fundamental incompatibilities. Our framework provides systematic validation mechanisms for deploying LLM-based automation in mission-critical telecommunications infrastructure.
\end{abstract}

\begin{IEEEkeywords}
Behavioral Personas, Generative Multi-Agent Systems, RAN Network Automation, O-RAN, Artificial Intelligence Safety Validation
\end{IEEEkeywords}

\section{Introduction}\label{sec:introduction}
\IEEEPARstart{T}{he} Open Radio Access Network (O-RAN) architecture disaggregates traditionally monolithic telecommunications systems into open, interoperable components, enabling intelligent automation through programmable applications (xApps, rApps) running on RAN Intelligent Controllers~\cite{oran_alliance_architecture,bonati2020intelligence}. However, realizing autonomous network management requires context-aware decision-making that balances competing objectives—energy efficiency, quality of service, and load distribution—in highly dynamic environments. Current automation approaches struggle with such multi-objective optimization, often resorting to conservative, one-size-fits-all solutions that fail to exploit O-RAN's full potential. The stakes for reliable automation are substantial: UK businesses lost £3.7 billion to internet failures in 2023~\cite{beaming2023}, while the European Union Agency for Cybersecurity documented significant telecommunications security incidents in 2024~\cite{naydenov2025}, highlighting the critical need for dependable automation in safety-critical infrastructure.

Large Language Models (LLMs) demonstrate remarkable capabilities in reasoning, planning, and code generation~\cite{gpt4_technical_report,brown2020language}, opening new possibilities for intelligent network automation. However, existing LLM-based multi-agent systems employ homogeneous decision-making strategies that cannot adapt to varying operational priorities~\cite{multiagent_network_control,llm_network_config}. Critically, they lack principled methods for customizing agent behavioral profiles and optimization preferences to specific network conditions or roles, and provide no systematic validation frameworks that assess both individual agent performance and emergent system-wide coordination dynamics before deployment in safety-critical infrastructure.

We address these limitations through a persona-driven multi-agent framework for autonomous O-RAN network management. \textit{Behavioral personas}—structured specifications encoding an agent's optimization priorities, risk tolerance, and decision-making style—systematically influence how five specialized agents (planner, coordinator, resource allocator, code generator, and analyzer) collaboratively approach network optimization tasks. To enable rigorous pre-deployment validation, we develop a \textit{decision-theoretic evaluation framework}~\cite{bell1988} measuring persona impact across three complementary dimensions: normative compliance (adherence to optimality criteria), prescriptive alignment (guideline consistency), and behavioral dynamics (emergent system properties). Comprehensive evaluation of 486 persona configurations reveals that customization yields highly heterogeneous impacts—performance deltas range from $+13.9$\% (well-aligned personas) to $-14.3$\% (misaligned personas)—with single-agent modifications propagating system-wide through cascading coordination effects.

This paper makes the following contributions:
\begin{enumerate}
    \item A persona-driven multi-agent architecture where behavioral personas influence decision-making across five specialized roles, evaluated across task complexity levels and retrieval architectures (RAG vs. GraphRAG).
    
    \item A three-dimensional evaluation framework: (a) \textit{normative evaluation} revealing Allocator scores collapse 19\% on multi-objective tasks while CodeAgent improves 3\%, demonstrating retrieval architecture effects; (b) \textit{prescriptive evaluation} uncovering failure signatures (0.50/0.50/0.50 patterns) indicating systematic persona rejection; (c) \textit{behavioral evaluation} capturing knowledge expansion, semantic convergence, and code quality progression.
    
   \item Evidence from 486 configurations that retrieval architecture fundamentally shapes behavior: GraphRAG agents show 27\% knowledge expansion improvement on complex tasks while RAG agents maintain 40\% convergence acceleration, revealing architectural constraints personas cannot overcome.

   \item Empirical demonstration of system-wide cascading effects: single persona changes propagate to all agents with system-wide deltas from $+6.21$\% to $-0.74$\%, and misalignments degrading 50\% of components despite local optimizations.

   \item Evidence that persona effectiveness depends on task structure: CodeAgent personas achieve 6\% differentiation on generation tasks while PlannerAgent personas show 2.9\% impact on planning, indicating role-specific customization limits.
 
   \item Open-source implementation including orchestration, persona system, three-dimensional evaluation pipeline, and all 486 configurations.
\end{enumerate}

Our findings reveal critical insights for safety-critical infrastructure deployment: (1) persona-agent alignment influences individual performance and coordination dynamics; (2) retrieval architecture constrains customization effectiveness—GraphRAG excels at multi-objective optimization but suffers semantic drift in negotiations while RAG provides stable convergence but limited expansion; (3) certain combinations exhibit fundamental incompatibilities detectable through evaluation (CreativeThinker-Planner: $-14.3$\%, FastFailAuditor abandonment); (4) aggressive local optimization degrades system-wide performance through cascading effects.

The remainder of this paper is organized as follows. Section~\ref{sec:relatedwork} reviews related work. Section~\ref{sec:architecture} presents system architecture and persona framework. Section~\ref{sec:evaluation} describes evaluation methodology. Section~\ref{sec:results} analyzes experimental results. Section~\ref{sec:conclusion} concludes.

\section{Related Work}\label{sec:relatedwork}

Generative AI has demonstrated potential in automating telecom tasks including code refactoring~\cite{du2023power} and network configuration~\cite{jiang2023large,xu2024cloudeval}. While recent work explores generative multi-agent systems (GMAS) for telecom~\cite{nezami2025generative,nezami2025descriptor,qin2025generative}, emphasis has been on benchmarking and automation rather than safety. To our knowledge, no research addresses safety risks in GMAS for telecom through systematic persona matching. We review four strands: (i) GMAS for telecom, (ii) safety challenges in GenAI, (iii) code quality and verification, and (iv) agent evaluation frameworks.

\textbf{Generative Multi-Agent Systems for Telecom.}
Panek et al.~\cite{t1} propose a modular framework for cloud-native 5G networks employing chain-of-thought reasoning. Zou et al.~\cite{t2} introduce edge-based multi-model deployment reducing centralized infrastructure reliance. Li et al.~\cite{t3} design multi-agent systems for agile telecom workflows with specialized agents handling prompt engineering and resource optimization. Wu et al.~\cite{autogen2023} introduce AutoGen for conversational multi-agent coordination. While these studies demonstrate GenAI orchestration feasibility in telecom, none systematically addresses safety risks or behavioral consistency.  Recently, Nezami et al.~\cite{nezami2025safety} proposed a safety evaluation framework for cooperative GMAS in telecom, focusing on miscoordination risks shaped by persona diversity, demonstrating progressive improvements in analyzer penalties and agent consistency across iterative runs. However, their work emphasizes safety pathways rather than systematic pre-deployment validation across normative, prescriptive, and behavioral dimensions, leaving a gap in comprehensive multi-dimensional assessment frameworks for mission-critical infrastructure deployment.

\textbf{Safety Challenges in GenAI Systems.}
Safety concerns span multiple dimensions. Freitas et al.~\cite{fp1} survey risks including hallucination and misinformation, auditing commercial chatbots. Wang et al.~\cite{t4} propose proactive multi-agent reinforcement learning, highlighting safety concerns including high dimensionality and non-stationarity. Bellay et al.~\cite{t5} outline risks including instability and incomplete contextual awareness. METR~\cite{metr2024} pioneered systematic assessment of catastrophic risks from autonomous AI, emphasizing independent evaluation before high-stakes deployment. In telecom, \cite{fp3} emphasize system-level safety and compositional explainability across heterogeneous models. These studies establish safety importance but focus on single-agent contexts or general MARL rather than mission-critical telecom infrastructure.

\textbf{Code Quality and Verification of LLM Outputs.}
LLM-generated code quality remains critical. Pearce et al.~\cite{9833571} found 40\% of GitHub Copilot code in security-critical scenarios contained vulnerabilities. Krebs and Mazumdar~\cite{10974840} found detailed prompts reduce errors per static-analysis metrics. Shaikhelislamov et al.~\cite{10899123} demonstrate static-analysis feedback reduces vulnerabilities by 20\%. Recent work integrates formal verification: Teuber and Beckert~\cite{11039275} couple LLMs with symbolic provers for Java verification, while Wang et al.~\cite{736074e14db24d6285027d84b443d60a} introduce SpecVerify achieving 46.5\% verification accuracy. Yang et al.~\cite{10438452} develop AVRE for automated verification of 5G protocol specifications. These highlight verification importance for infrastructure-critical domains. In telecom, code generation must satisfy domain-specific safety requirements beyond general quality metrics, including O-RAN constraints~\cite{oran_alliance_specs} for resource allocation, latency bounds, and failover behavior.

\textbf{Agent Evaluation Frameworks.}
Wang et al.~\cite{wang2024towards} propose benchmarks for agents' social interaction through action-level metrics. Samuel et al.~\cite{samuel2024personagym} introduce PersonaGym with human-aligned metrics for persona adherence, though focused on conversational personas rather than operational personas encoding domain-specific optimization strategies. Mao et al.~\cite{mao2023alympics} present simulation frameworks for game-theoretic scenarios. The UK AI Security Institute developed Inspect AI~\cite{inspectai2024} for large-scale capability and safety evaluation of AI system across standardized benchmarks. Recent work explores LLMs as evaluators: Zheng et al.~\cite{zheng2023judging} demonstrate GPT-4 as scalable judge for multi-turn conversations, while Liu et al.~\cite{liu2023g} propose G-Eval correlating with human judgments, though focusing on general-domain evaluation without validation for mission-critical infrastructure. In telecom, NIKA~\cite{nika2024} provides network arenas for troubleshooting benchmarks, while GSMA~\cite{gsma2024} developed comprehensive evaluation suites (TeleQnA, TeleLogs, TeleMATH, TeleYAML) for single-agent telecom tasks. While advancing language-agent evaluation and establishing telecom benchmarks, emphasis remains on single-agent completion or general capability rather than multi-agent safety, behavioral compliance, and coordination reliability.

Prior work establishes GMAS feasibility in telecom and highlights safety challenges, but lacks systematic frameworks validating whether LLM-driven multi-agent systems exhibit reliable, safety-compliant behavior in mission-critical infrastructure. Existing evaluation focuses on single-agent task completion, not whether agents produce persona-optimal outputs (normative compliance), maintain prescribed behaviors (prescriptive alignment), and sustain healthy coordination (behavioral dynamics). Our work addresses this gap through decision-theoretic three-dimensional evaluation, enabling systematic pre-deployment validation of persona-driven multi-agent systems for O-RAN automation.

\section{System Architecture and Operation}\label{sec:architecture}
We present a persona-driven multi-agent system for autonomous O-RAN network management where five specialized agents collaborate through a structured workflow to transform high-level optimization objectives into validated, executable solutions.

\begin{figure*}
    \centering
    \includegraphics[clip, trim=0cm 3.5cm 0cm 7cm,width=\textwidth]{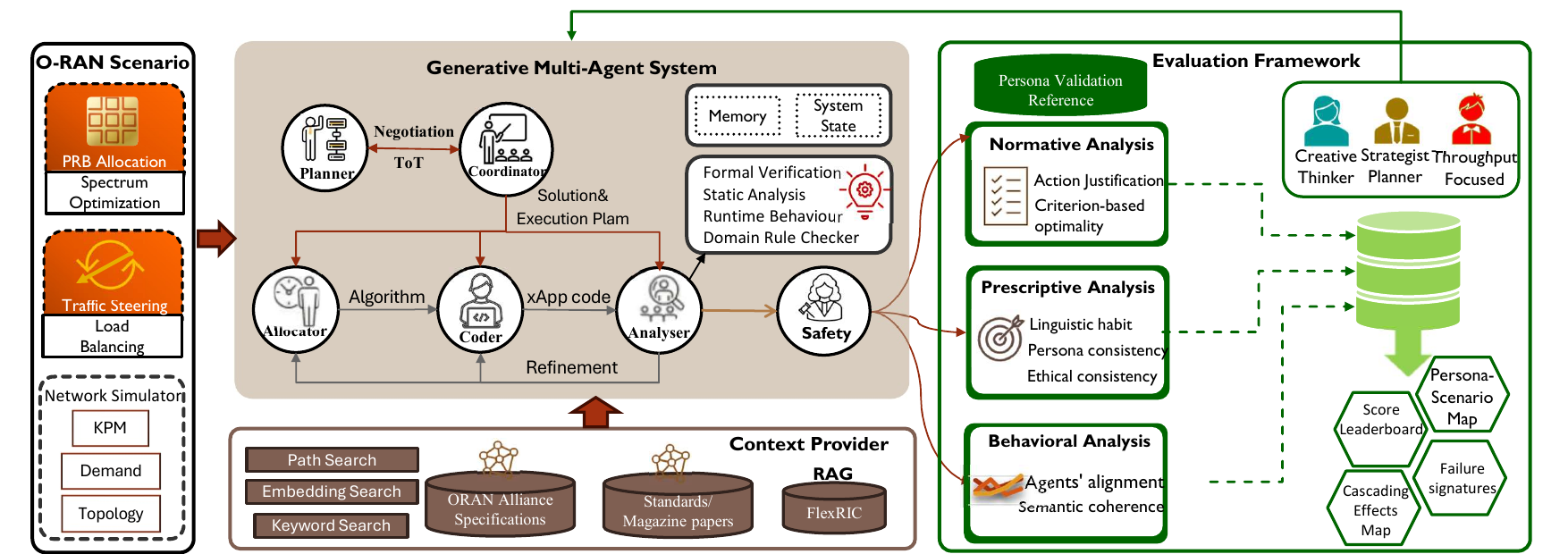}
    \caption{Multi-agent system architecture showing orchestration, execution, and validation subsystems with knowledge layer integration.}
    \label{fig:architecture}
\end{figure*}

Network management involves inherent trade-offs between conflicting objectives—throughput versus energy efficiency, latency versus reliability. The optimal balance varies by operational context: emergencies require risk-averse strategies with conservative resource allocation, while peak traffic events justify aggressive throughput optimization. However, existing automation systems apply fixed strategies regardless of context, network state, or operational priorities. This motivates a framework where agents adopt configurable behavioral personas encoding operational philosophies, optimization preferences, and risk tolerances that systematically influence decision-making throughout solution generation.

\subsection{Multi-Agent System}
The system operates through a coordinated multi-phase workflow (Fig.~\ref{fig:architecture}) wherein PlannerAgent explores multiple candidate solutions, followed by negotiation and selection with CoordinatorAgent. The chosen plan is translated into algorithmic specifications by ResourceAllocationAgent, implemented as executable code by CodeAgent, and validated across multiple dimensions by AnalyserAgent. Failed validations trigger targeted refinement cycles, ensuring convergence toward deployable, policy-compliant solutions.

The architecture comprises three functional subsystems supported by domain knowledge integration and episodic memory. When the system receives a network optimization query (e.g., ``optimize PRB allocation for QoS differentiation''), the orchestration subsystem initiates the workflow. While this architecture enables sophisticated solution generation, its safety and reliability in mission-critical infrastructure cannot be assumed. Section~\ref{sec:evaluation} presents our three-dimensional evaluation framework that systematically validates whether persona-driven agents produce optimal outputs (normative), maintain prescribed behaviors (prescriptive), and coordinate effectively (behavioral) before deployment.

\subsubsection{Orchestration Subsystem}
The orchestration subsystem comprises two agents responsible for solution exploration and execution planning:

\textbf{PlannerAgent} explores multiple solution approaches using Tree-of-Thoughts (ToT) reasoning~\cite{yao2023tree}, which extends standard LLM reasoning by generating diverse solution paths, evaluating intermediate steps, and maintaining multiple candidate trajectories. It generates multiple candidate paths (typically three), each reflecting a distinct strategic approach. These trajectories are evaluated against feasibility constraints and optimization objectives, and iteratively refined through multi-round negotiation with CoordinatorAgent. The agent leverages GraphRAG to retrieve relevant O-RAN specifications and prior research, ensuring solutions remain technically grounded and standards-compliant. ToT enables self-correction and exploration of diverse optimization strategies, supporting persona-driven trade-offs under uncertain network conditions.

\textbf{CoordinatorAgent} manages the overall execution lifecycle by selecting the most promising solution path from PlannerAgent's proposals and orchestrating downstream agents. It negotiates with PlannerAgent across multiple rounds to reach alignment on the solution approach, selects the most appropriate path based on feasibility and expected performance, and decomposes the strategy into discrete execution steps. It then orchestrates execution subsystem agents and manages iterative refinement when quality issues are detected. The negotiation protocol ensures persona-driven strategic intent aligns with operational constraints before execution begins.

\subsubsection{Execution Subsystem}

The execution subsystem translates high-level plans into concrete implementations through two specialized agents:

\textbf{ResourceAllocationAgent} generates resource allocation strategies as structured pseudocode specifications. Given CoordinatorAgent's execution plan, it designs algorithms accounting for QoS requirements, load balancing constraints, and current network state. It produces detailed pseudocode with explicit variable definitions, control-flow logic, and algorithmic steps. To ensure standards compliance, the agent retrieves relevant O-RAN specifications~\cite{oran_alliance_specs} through GraphRAG and provides algorithmic rationales explaining key design decisions and optimization trade-offs. The pseudocode serves as an intermediate representation enabling validation before code generation and facilitating human oversight of algorithmic decisions.

\textbf{CodeAgent} implements executable code faithful to ResourceAllocationAgent's pseudocode specification. It translates the pseudocode into production-quality Python or C++ implementations, retrieving established implementation patterns from the FlexRIC codebase\footnote{FlexRIC: \url{https://gitlab.eurecom.fr/mosaic5g/flexric}} via RAG. Generated code incorporates defensive programming practices, including error handling and type safety, accompanied by inline documentation. The agent maintains strict adherence to the algorithmic specification to ensure consistency between design intent and executable behavior. Separation between algorithmic design and implementation enables independent validation of logic and code quality, essential for safety-critical telecom infrastructure.

\subsubsection{Validation and Refinement Governance Subsystem}

\textbf{AnalyserAgent} performs system-level analysis and validation across the entire solution generation pipeline prior to deployment, ensuring consistency between high-level intent, algorithmic design, and executable code. Rather than evaluating generated code in isolation, it jointly analyzes outputs of upstream agents, including candidate solution strategies from PlannerAgent, execution plans from CoordinatorAgent, and pseudocode from ResourceAllocationAgent, together with current O-RAN network state and demand context.

Validation is conducted along four complementary dimensions. Structural quality is assessed using static analysis tools to evaluate readability, modularity, and coding standards. Runtime behavior is examined through sandboxed execution to identify execution failures, unhandled exceptions, and performance bottlenecks. Policy compliance is enforced via abstract syntax tree analysis to detect prohibited API usage, unsafe function calls, and security violations. Lightweight formal checks verify properties such as type safety, exception handling completeness, and deterministic execution patterns.

Based on aggregated validation results, AnalyserAgent determines whether refinement is required and identifies the appropriate agent to address detected issues, distinguishing between algorithmic flaws and implementation defects. It provides targeted feedback to either ResourceAllocationAgent or CodeAgent, triggering a bounded refinement loop. This loop continues until all quality dimensions satisfy predefined thresholds or a maximum iteration limit is reached, ensuring controlled convergence toward deployable, policy-compliant artifacts.

\subsection{Persona Integration}

Personas~\cite{wang2024unleashing,samuel2024personagym} characterize each agent's behavioral traits and optimization priorities, shaping both decision-making processes and communication patterns. Unlike conversational personas designed for user interaction, these personas influence agents' underlying reasoning strategies and optimization objectives. Personas are automatically generated by LLMs, guided by structured specifications encompassing: (a) normative criteria specifying optimization goals that constitute ``optimal'' behavior for an agent's role (e.g., maximizing throughput, minimizing latency, ensuring fairness); (b) prescriptive guidelines covering linguistic style, consistency, and ethical considerations governing how the agent operates; and (c) strategic orientation defining high-level decision-making priorities, such as risk aversion versus performance optimization, or minimalist versus comprehensive approaches.

Representative personas include: \textit{RiskAverse Planner} (avoids high-risk or uncertain plan steps, favoring predictable sequences), \textit{ThroughputFocused Coordinator} (prioritizes system throughput as the main planning objective), \textit{ThroughputMaximizer ResourceAllocation} (targets peak bandwidth utilization and aggregate network capacity), \textit{Minimalist Coder} (prefers clean, minimal, maintainable code), and \textit{StrictAssessor Analyser} (enforces rigorous standards for structure, behavior, policy compliance, and formal correctness, prioritizing enterprise-grade safety).

Prior research shows that role-specific personas can enhance task-specific reasoning compared to generic prompting~\cite{samuel2024personagym,mao2023alympics}. We systematically evaluate combinations of these and additional personas across all five agents, testing 486 configurations to assess how persona diversity affects solution quality and system reliability (Section~\ref{sec:results}).

\subsection{Knowledge Layer}
Agent reasoning leverages domain-specific knowledge through complementary retrieval mechanisms tailored to each agent's role. PlannerAgent combines two GraphRAG retrieval methods: path-based traversal of a telecom knowledge graph capturing problem-solution relationships, and embedding-based semantic search over the same graph. This dual retrieval provides rich structured context informing multi-path solution exploration. ResourceAllocationAgent employs embedding-based GraphRAG retrieval to obtain O-RAN specification context relevant to resource allocation, supporting precise algorithmic design grounded in network demands and regulatory constraints. CodeAgent relies on RAG to retrieve unstructured documents such as FlexRIC implementation examples, code patterns, and technical reports, assisting in generating robust, production-quality code consistent with best practices. All retrieval outputs are formatted via unified context provider interfaces, allowing agents to seamlessly integrate structured and unstructured knowledge into their reasoning and generation workflows.

\subsection{Memory and State Management}

The system incorporates episodic memory to maintain history of each agent's decision-making process, enabling agents to recall how similar problems were previously addressed, avoid repeating failed strategies from past refinement cycles, build upon successful patterns from earlier tasks, and maintain consistency throughout multi-round negotiations and refinements. A Memory Manager stores per-agent trajectories, including inputs, intermediate reasoning steps, final outputs, and refinement feedback, allowing agents to operate with continuity rather than treating each invocation as an isolated event.


\subsection{Extensibility and Deployment}

The architecture supports modularity and extensibility across development, evaluation, and production environments. New domain-specific agents can be integrated by implementing the standardized agent interface. Additional validation dimensions may be incorporated through plugin-based checkers, and custom personas can be defined through structured specification files. The knowledge layer permits addition of new retrieval corpora or knowledge graphs without code changes.

The framework supports autonomous execution of solution workflows for evaluation purposes, enabling systematic testing across diverse network scenarios. Deployment modes include local sandboxes for development and testing, laboratory testbeds for KPI-driven validation using simulated networks, and production deployments enforcing strict policy compliance, human-in-the-loop oversight, and real-time monitoring. Governance parameters (quality thresholds, refinement limits, policy constraints) are configurable per mode, enabling tailored autonomy-safety trade-offs.

\section{Evaluation Framework}\label{sec:evaluation}
We introduce an offline evaluation framework to analyze and quantify persona-driven agent behavior for research and benchmarking. Unlike the Validation and Refinement Governance Subsystem, which operates within the system to ensure pre-deployment correctness, this evaluation framework is applied externally, functioning independently of runtime decision-making. The framework is grounded in decision theory, comprising three complementary branches: normative evaluation (optimality assessment), prescriptive evaluation (behavioral guideline adherence), and behavioral evaluation (emergent system dynamics). 

\subsection{Normative Evaluation}

Normative decision theory establishes criteria for optimal decisions by rational actors~\cite{bell1988}. This branch addresses: \textit{What is the optimal output for a rational agent with specific persona characteristics in a given environment?}

\textbf{Problem Formulation.} Let $\mathcal{A} = \{a_1, a_2, \ldots, a_n\}$ denote the set of agents in our multi-agent system, and $\mathcal{P} = \{p_1, p_2, \ldots, p_m\}$ represent the set of available personas. A persona assignment function $\psi: \mathcal{A} \rightarrow \mathcal{P}$ maps each agent to a persona. Given an agent $a_i$ with assigned persona $p_j = \psi(a_i)$ and task input $x$, the agent produces output $o = a_i(x; p_j)$.

We conduct normative evaluation through \textbf{criterion-based optimality assessment}, where each persona is defined by a set of optimality criteria $\mathcal{O}_{p_j} = \{o_1, o_2, \ldots, o_k\}$, characterizing optimal behavior for that role. For example, an ``EfficiencyMaximizer'' resource allocator is defined by criteria such as ``maximizes aggregate throughput'' and ``optimizes spectral efficiency.'' Formally, we evaluate a normative alignment function:
\begin{equation}
N(o, p_j) = \frac{1}{|\mathcal{O}_{p_j}|} \sum_{c \in \mathcal{O}_{p_j}} s(o, c)
\end{equation}

\noindent where $s(o, c) \in [0,1]$ evaluates the degree to which output $o$ satisfies optimality criterion $c$.

\textbf{Criterion-Based Assessment.} Unlike the Expected Action task methodology \cite{samuel2024personagym} which presents explicit action options and evaluates whether agents select the mathematically optimal action, our approach evaluates normative alignment through LLM-based criterion satisfaction assessment. For each agent output, an LLM judge evaluates how well the output satisfies each normative criterion, providing scores on a continuous [0,1] scale with textual justifications.

This criterion-based approach enables normative evaluation of complex, open-ended outputs where mathematical utility calculation would be infeasible. While differing methodologically from Expected Action evaluation, it addresses the same fundamental question: does the agent's output align with the optimality standards prescribed by its persona?

\textbf{Implementation.} For each evaluation, we construct a prompt providing the LLM judge with: (a) the persona's normative criteria, (b) the agent's generated output and reasoning, and (c) relevant context from upstream agents. The judge returns structured assessments including per-criterion scores, justifications grounded in specific evidence, and an aggregate normative alignment score. This balances the contextual understanding capabilities of LLMs \cite{brown2020,wei2022} with structured evaluation requirements for reproducibility.


\subsection{Prescriptive Evaluation}

Prescriptive decision theory provides guidelines for how agents should behave 
within cognitive and environmental constraints \cite{bell1988,simon1955}. This branch addresses: \textit{How should an agent with specific characteristics behave in a given environment?} We implement prescriptive evaluation through three interconnected dimensions, each capturing distinct aspects of guideline adherence.

\textbf{Definition 1 (Linguistic Habits).} For persona $p_j$, let 
$\mathcal{L}_{p_j} = \{\ell_1, \ell_2, \ldots, \ell_5\}$ denote five linguistic prescriptions characterizing how the agent should communicate. The linguistic compliance score is:

\begin{equation}
\Pi_L(o, p_j) = \frac{1}{|\mathcal{L}_{p_j}|} \sum_{\ell \in \mathcal{L}_{p_j}} s(o, \ell)
\end{equation}

\noindent where $s(o, \ell) \in [0,1]$ evaluates adherence to linguistic 
prescription $\ell$.

\textbf{Definition 2 (Persona Consistency).} Let $\mathcal{C}_{p_j} = \{c_1, c_2, \ldots, c_5\}$ represent behavioral consistency prescriptions for persona $p_j$. The consistency compliance score is:

\begin{equation}
\Pi_C(o, p_j) = \frac{1}{|\mathcal{C}_{p_j}|} \sum_{c \in \mathcal{C}_{p_j}} s(o, c)
\end{equation}

\textbf{Definition 3 (Ethical Compliance).} Let $\mathcal{E}_{p_j} = \{e_1, e_2, \ldots, e_5\}$ denote ethical conduct prescriptions. The ethical compliance score is:

\begin{equation}
\Pi_E(o, p_j) = \frac{1}{|\mathcal{E}_{p_j}|} \sum_{e \in \mathcal{E}_{p_j}} s(o, e)
\end{equation}

\textbf{Overall Prescriptive Score.} The aggregate prescriptive compliance is:

\begin{equation}
\Pi(o, p_j) = \omega_L \cdot \Pi_L(o, p_j) + \omega_C \cdot \Pi_C(o, p_j) + \omega_E \cdot \Pi_E(o, p_j)
\end{equation}

\noindent where $\omega_L, \omega_C, \omega_E \geq 0$ and $\omega_L + \omega_C + \omega_E = 1$ are dimension weights. We use equal weights ($\omega_L = \omega_C = \omega_E = 1/3$).

\textbf{Linguistic Habits.} This dimension assesses whether agents' communication patterns align with prescriptive expectations for their assigned persona~\cite{samuel2024personagym}. Drawing from computational sociolinguistics~\cite{danescu2013} and dialogue systems~\cite{li2016}, it evaluates: (a) lexical choices and domain-specific jargon, (b) syntactic structures and sentence complexity, (c) discourse markers and pragmatic conventions, (d) formality level and register consistency, and (e) affective tone and hedging strategies.

For example, a ``Strategist'' coordinator uses forward-looking temporal language (mentioning long-term impacts like scalability), maintain formal register avoiding colloquialisms, and reference system-wide patterns rather than isolated incidents. A ``Tactician'' uses action-oriented imperatives with urgency-aware wording (e.g., respond, adapt now), employ present-tense constructions referencing immediate conditions, and focus on actionable changes. These prescriptive patterns enable evaluation of persona-specific linguistic alignment.

\textbf{Persona Consistency.} This dimension evaluates behavioral consistency with prescribed persona attributes. Grounded in agent architecture theory \cite{bratman1987,rao1995} and organizational role theory \cite{biddle1986}, it assesses whether agents maintain characteristic behaviors across diverse scenarios.

A ``RiskAverseCoordinator'' is prescribed to: never propose experimental approaches, validate each step for known reliability, prefer predictable methods even if slower, avoid paths with weak evidential support, and maintain conservative safety margins. A ``MinimalistCoordinator'' is prescribed to: minimize plan steps, use fewest agents necessary, remove unnecessary actions, prioritize simplicity, and create compact plans. Violations—such as a MinimalistCoordinator producing elaborate 
multi-step plans—indicate prescriptive non-compliance.

\textbf{Ethical Compliance.} This dimension ensures agents follow prescriptive guidelines for responsible conduct. Drawing from trustworthy AI frameworks \cite{dignum2019,floridi2018} and agent accountability research \cite{cointe2016}, it evaluates: (a) transparency in capability representation and uncertainty acknowledgment, (b) respect for data confidentiality and privacy, (c) adherence to authority boundaries, (d) fairness and bias avoidance, and (e) responsible handling of conflicting objectives.

For example, a ``ThroughputMaximizer'' is prescribed to acknowledge when pursuing maximum throughput conflicts with fairness objectives, transparently report when network conditions make optimization predictions uncertain, and avoid overclaiming achievable performance gains. A ``RiskAverseCoordinator'' is prescribed to explicitly state safety margins in resource allocation, acknowledge data limitations when making conservative decisions, and transparently explain why certain 
experimental approaches are avoided.

In network management contexts, ethical prescriptions include: avoiding overclaiming optimization capabilities, transparently reporting prediction uncertainties, maintaining caution in safety-critical decisions, acknowledging constraints in available data, and using transparent rationale when prioritizing objectives. These prescriptions are crucial as agent decisions directly impact service quality and system reliability.

\subsection{Behavioral Evaluation}

While normative and prescriptive evaluation assess individual agent outputs against established standards, behavioral evaluation examines emergent system-level dynamics arising from multi-agent interaction. Drawing from distributed systems theory~\cite{coulouris2011} and organizational cybernetics~\cite{beer1972}, this branch addresses: \textit{How do agents' behavioral patterns evolve and interact within the multi-agent system?} We implement behavioral evaluation through four complementary dimensions capturing different aspects of emergent agent behavior and system dynamics.

\textbf{Definition 4 (Semantic Convergence).} Let $\mathcal{E}(c_i)$ denote the embedding representation of context $c_i$ retrieved or generated by agent $a_i$. For iterative refinement processes, we track how contexts evolve across runs. Let $c_i^{(t)}$ denote the context at iteration $t$. The semantic distance between consecutive iterations is:

\begin{equation}
D_{sem}^{(t)} = 1 - \frac{\mathcal{E}(c_i^{(t)}) \cdot \mathcal{E}(c_i^{(t+1)})}{\|\mathcal{E}(c_i^{(t)})\| \cdot \|\mathcal{E}(c_i^{(t+1)})\|}
\end{equation}

\noindent where $\cdot$ denotes dot product and $\|\cdot\|$ denotes Euclidean norm. We measure semantic convergence across refinement runs to quantify whether agent outputs stabilize toward consistent solutions or continue to explore diverse alternatives. For PlannerAgent with multi-turn negotiation, we track distance between embedding context (retrieved domain knowledge) and problem-solution context (evolving negotiation state) to measure contextual alignment degradation.

\textbf{Definition 5 (Knowledge Expansion).} Agents augment user-provided concepts with domain knowledge from their contexts. Let $\mathcal{V}_{user}$ denote the concept vocabulary from user input, and $\mathcal{V}_{agent}(a_i)$ denote concepts introduced by agent $a_i$. The novel concepts are:

\begin{equation}
\mathcal{V}_{novel}(a_i) = \mathcal{V}_{agent}(a_i) \setminus \mathcal{V}_{user}
\end{equation}

We classify novel concepts using domain taxonomy $\mathcal{T} = \{T_{\text{networking}}, T_{\text{resource}}, T_{\text{qos}}, T_{\text{ml}}, T_{\text{opt}}, T_{\text{oran}}, T_{\text{traffic}}, T_{\text{energy}}, T_{\text{prog}}\}$ covering nine technical domains: networking protocols, resource management, QoS performance, machine learning, optimization algorithms, O-RAN architecture, traffic demand, energy efficiency, and programming constructs. The expansion and drift metrics are:

\begin{equation}
\begin{aligned}
\mathcal{V}_{expansion}(a_i) &= \{v \in \mathcal{V}_{novel}(a_i) : \exists T_j \in \mathcal{T}, v \in T_j\} \\
\mathcal{V}_{drift}(a_i) &= \{v \in \mathcal{V}_{novel}(a_i) : \forall T_j \in \mathcal{T}, v \notin T_j\}
\end{aligned}
\end{equation}

The expansion and drift ratios quantify knowledge contribution patterns:

\begin{equation}
r_{exp}(a_i) = \frac{|\mathcal{V}_{expansion}(a_i)|}{|\mathcal{V}_{novel}(a_i)|}, \quad r_{drift}(a_i) = 1 - r_{exp}(a_i)
\end{equation}

High expansion ratios indicate effective domain knowledge integration, while high drift ratios suggest agents introduce off-topic concepts that may indicate retrieval misalignment or hallucination.

\textbf{Definition 6 (Code Quality Dynamics).} For code generation pipeline, we evaluate quality evolution across refinement iterations. Quality is assessed across four equally-weighted dimensions (25\% each):

\begin{equation}
Q_{code} = \frac{1}{4}(Q_{structure} + Q_{runtime} + Q_{policy} + Q_{guarantees})
\end{equation}

\noindent where $Q_{structure}$ measures code organization, readability, and modularity; $Q_{runtime}$ measures execution efficiency and runtime behavior; $Q_{policy}$ measures O-RAN specification compliance and approved function usage; and $Q_{guarantees}$ measures type safety, error handling, and formal verification requirements.

For iterative refinement, we track quality evolution across runs. Let $Q^{(t)}$ denote quality at iteration $t$. The quality improvement is:

\begin{equation}
\Delta_{quality}^{(t)} = Q^{(t+1)} - Q^{(t)}
\end{equation}

Tracking component-specific trajectories reveals which quality dimensions benefit from multi-agent refinement and which remain bottlenecks requiring specialized attention.

\section{Experimental Setup}\label{sec:experiments}

We evaluate our persona-driven multi-agent system on two representative O-RAN automation tasks that exercise different agent capabilities: PRB (Physical Resource Block) allocation, which involves single-objective resource optimization requiring domain knowledge of radio resource management, and traffic steering, which requires multi-objective balancing across network elements. Both tasks involve real-time decision-making under operational constraints where agents must maintain prescribed 
behavioral roles despite complexity.

We systematically test persona combinations across all five agents. Each agent is configured with one of three distinct personas tailored to its role, available on \url{https://github.com/cheddarhub/MultiAgentSafetyFramework}. This yields $3^5 = 243$ unique persona combinations. We evaluate each combination on both tasks (PRB allocation and traffic steering), with multiple refinement iterations per configuration, resulting in 486 total configuration-task evaluations. All agents are powered by GPT-4o mini (gpt-4o-mini-2024-07-18), accessed via the OpenAI API with temperature set to 0.7 for generation tasks and 0.0 for evaluation tasks to ensure reproducibility. The negotiation round and refinement loop are bounded at a maximum of 3 iterations per configuration.

For each configuration-task pair, we apply the three-dimensional evaluation framework (Section~\ref{sec:evaluation}) using GPT-4o mini as the LLM judge. Normative evaluation scores persona-specific criterion satisfaction on [0,1] scales. Prescriptive evaluation assesses linguistic habits, persona consistency, and ethical compliance through structured guideline adherence. Behavioral evaluation tracks coordination efficiency, knowledge expansion, and decision consistency. All evaluations use structured prompts providing persona criteria, agent outputs, and upstream context, with judges returning quantitative scores and textual justifications.


\begin{figure*}
    \centering
    \includegraphics[clip, trim=0cm 1.5cm 0cm 3.7cm,width=\linewidth]{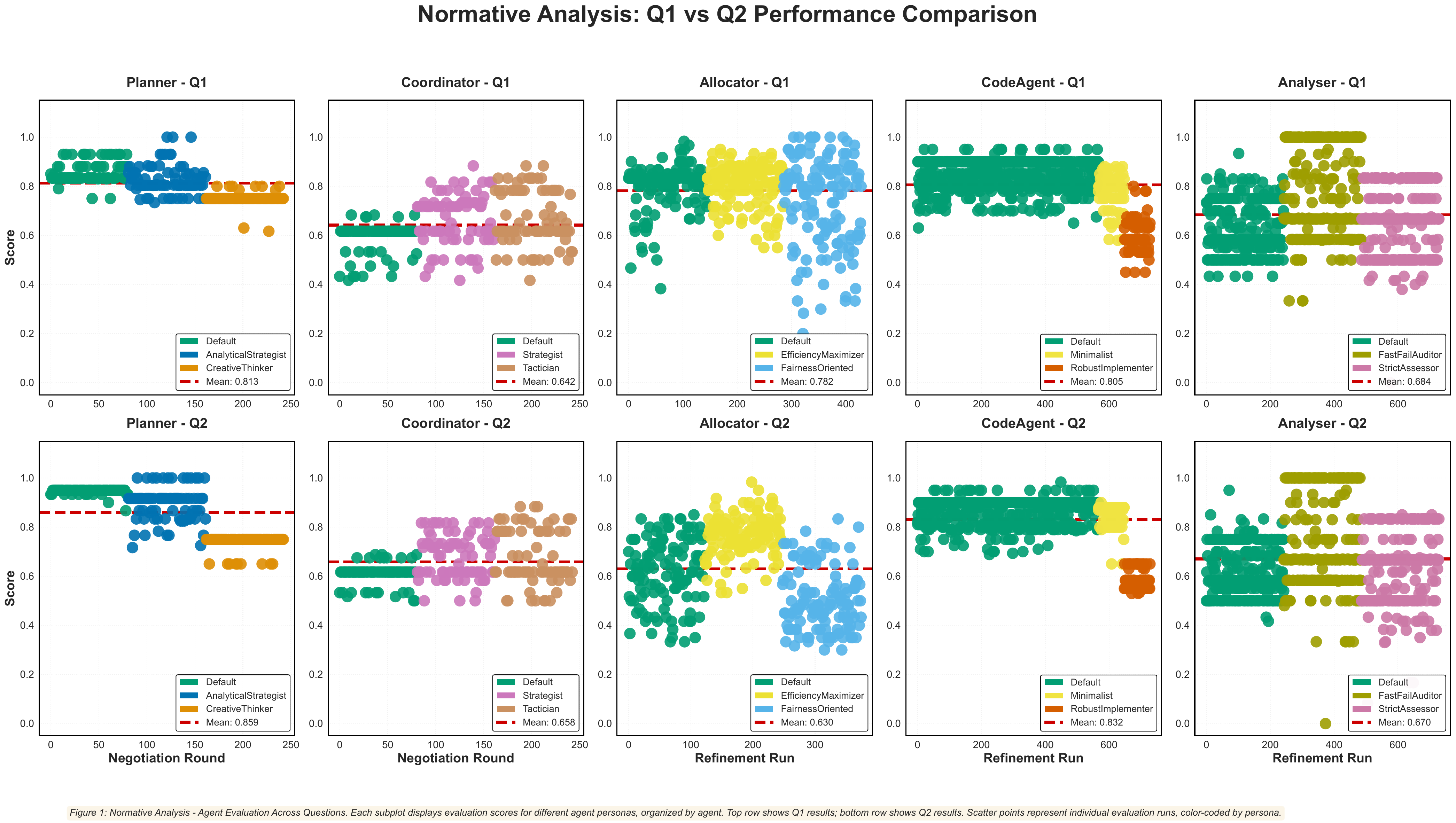}
    \caption{Normative Analysis - Agent Evaluation Across Questions. 
Each subplot displays evaluation scores for different agent personas, 
organized by agent. Top row shows Q1 results; bottom row shows Q2 results. 
Scatter points represent individual evaluation runs, color-coded by persona.}
    \label{fig:normative}
\end{figure*}

\section{Results and Discussion}\label{sec:results}
We evaluate our framework on two O-RAN optimization challenges (supplementary material): \textbf{Q1 (PRB Allocation)} minimizes energy under demand uncertainty ($\pm 20$\%) and QoS constraints (5 Mbps/user, RSRP $\geq -110$ dBm); \textbf{Q2 (Handover Optimization)} balances network load through strategic user reassignments while minimizing handover overhead. Both enforce identical utilization targets (60--80\%, max 90\%) and QoS thresholds, enabling controlled persona comparison. Supplementary material summarizes persona configuration for Planner agent, comprising cognitive style description, normative criteria, and prescriptive criteria. Complete specifications appear our GitHub repository\footnote{\url{https://github.com/cheddarhub/MultiAgentSafetyFramework}}.



\subsection{Normative Analysis}
Normative evaluation reveals task-dependent agent performance hierarchies. On Q1 (single-objective), Planner and CodeAgent achieve highest normative alignment ($\mu = 0.81$), followed by Allocator ($\mu = 0.78$), Analyser ($\mu = 0.68$), and Coordinator ($\mu = 0.64$). On Q2 (multi-objective), this hierarchy shifts: Planner and CodeAgent improve modestly (3\% each), while Allocator degrades substantially (19\% decline to $\mu = 0.63$), indicating multi-objective scenarios challenge criterion satisfaction. CodeAgent maintains low variance (std $\approx 0.10$) across tasks, while Allocator exhibits highest variance (std $= 0.13$--$0.16$).


Persona effects vary by agent and task complexity. Allocator shows minimal persona differentiation on Q1 (9\% span) but pronounced effects on Q2: EfficiencyMaximizer ($\mu = 0.77$) achieves 53\% higher scores than FairnessOriented ($\mu = 0.50$), revealing fairness prescriptions poorly satisfy optimality criteria in multi-objective contexts. CodeAgent exhibits consistent persona patterns: Default achieves highest scores, while RobustImplementer scores 28--32\% lower across tasks, suggesting defensive programming compromises optimality criterion satisfaction. Analyser's FastFailAuditor maintains 24\% advantage over Default and StrictAssessor. Planner's Default persona achieves exceptional scores ($\mu_{Q2} = 0.92$--$0.95$), outperforming specialized personas (AnalyticalStrategist, CreativeThinker) by 19--22\%.


These patterns reveal architectural constraints on normative alignment. Allocator's Q2 collapse—particularly for FairnessOriented (50\% drop)—demonstrates that while GraphRAG successfully retrieves O-RAN specifications, it fails to synthesize them into solutions satisfying multi-objective optimality criteria. Fairness prescriptions fundamentally conflict with efficiency optimization in resource-constrained scenarios. Conversely, CodeAgent's Q1-to-Q2 improvement reflects RAG's provision of concrete FlexRIC implementation patterns that satisfy code quality criteria independent of problem complexity. These divergent trajectories indicate retrieval architecture fundamentally constrains criterion satisfaction: GraphRAG excels at knowledge breadth but struggles with synthesis; RAG provides stable templates but limited flexibility. Within-agent persona variance (up to 53\%) suggests complex interactions between persona prescriptions, task characteristics, and retrieval mechanisms. Frontier LLMs demonstrate sufficient O-RAN knowledge for single-objective tasks but insufficient multi-objective reasoning, revealing critical deployment limitations without systematic validation.

\subsection{Prescriptive Analysis}

\begin{figure*}
    \centering
    \includegraphics[clip, trim=0cm 0cm 0cm 0cm,width=\linewidth]{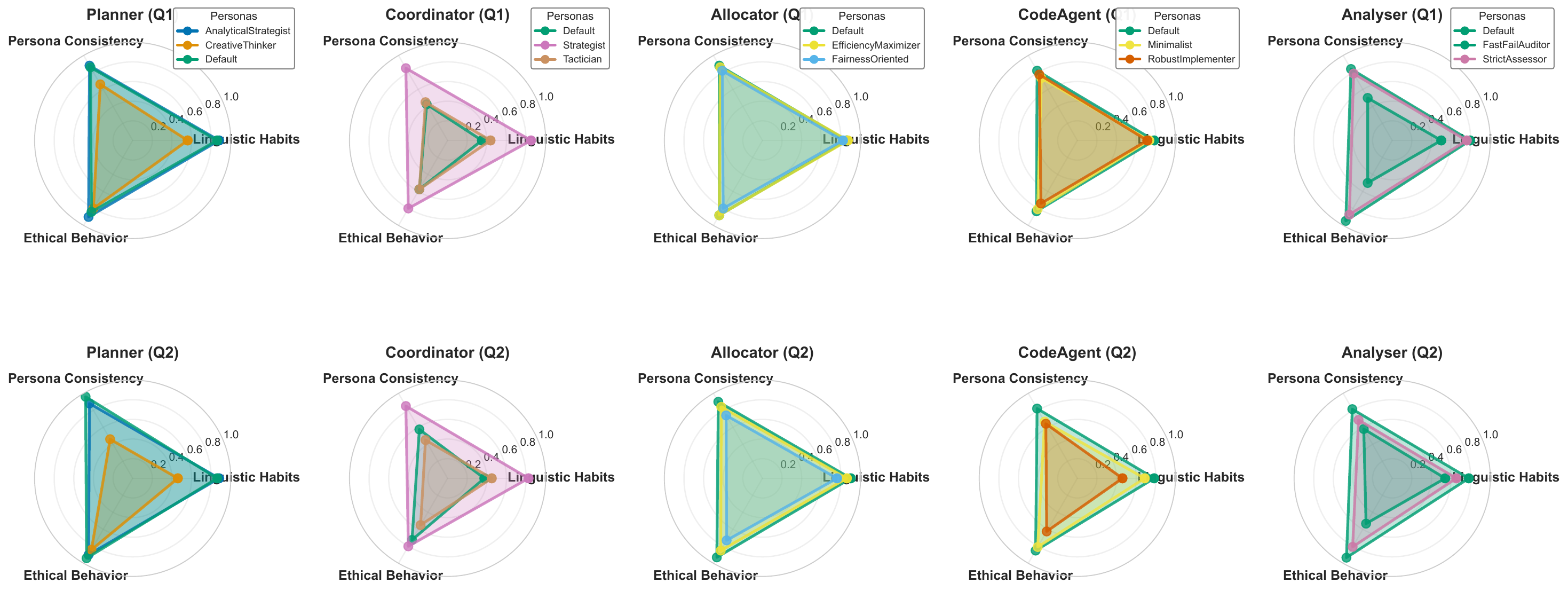}
    \caption{Prescriptive evaluation across Linguistic Habits, Persona Consistency, and Ethical Behavior for five agents under Q1 (top) and Q2 (bottom). Three personas per agent (colored triangles).}
    \label{fig:prescriptive}
\end{figure*}

Prescriptive evaluation (Fig.~\ref{fig:prescriptive}) reveals substantial guideline adherence variation (0.45--0.92 range). High-performing configurations demonstrate strong compliance scores: Planner-Default (0.85--0.92), Allocator-Default (0.87--0.91), and Coordinator-Strategist (0.83 consistently). Critical failure patterns emerge: Analyser-FastFailAuditor exhibits a distinctive 0.50/0.50/0.50 signature across all dimensions on Q1, indicating complete persona abandonment. Coordinator-Default scores 0.45 overall on Q1 (improving to 0.56 on Q2) with severe Linguistic Habits deficits (0.35 and 0.38 respectively), while Coordinator-Tactician achieves only 0.49 overall on both tasks. Individual dimensions show greater disparities: Linguistic Habits (0.35--0.90), Persona Consistency (0.43--0.96), and Ethical Behavior (0.53--0.95). Coordinator-Strategist maintains 0.83 compliance, a 71\% advantage over Default and 84\% over Tactician, demonstrating configuration-dependent adherence within identical architectures.


Dimensional analysis reveals distinct compliance patterns. Ethical Behavior achieves highest scores (0.80--0.95 typical range), with agents reliably adhering to transparency, capability acknowledgment, and responsible conduct prescriptions. Even poor-performing configurations maintain strong ethical adherence: Coordinator-Default scores 0.58--0.73 on Ethical Behavior despite 0.45--0.56 overall; Coordinator-Tactician achieves 0.58 ethical compliance while scoring 0.49 overall. Ethical guidelines prove more robust to persona variations than linguistic or behavioral prescriptions. Persona Consistency exhibits moderate variability (0.42--0.96), with most agents achieving 0.75--0.90 compliance, indicating reasonable maintenance of prescribed behavioral patterns. However, specific personas struggle: Planner-CreativeThinker drops from 0.66 to 0.46 (30\% decline), and Coordinator-Tactician maintains only 0.45 consistency. Linguistic Habits demonstrates highest variability and lowest scores (0.35--0.90), revealing persona-specific communication patterns are most challenging to maintain. Coordinator-Default achieves only 0.35--0.38, Planner-CreativeThinker declines from 0.56 to 0.46, and CodeAgent-RobustImplementer drops from 0.71 to 0.46 (35\% decline). Current LLM-based agents readily adopt ethical guidelines and maintain moderate behavioral consistency but struggle with prescribed linguistic patterns—domain jargon, syntactic structures, and formality registers.

Cross-task compliance trajectories illuminate task complexity effects. Allocator improves across all dimensions (0.87 $\rightarrow$ 0.91 overall, 5\% gain), with balanced improvements in Linguistic Habits (0.86 $\rightarrow$ 0.90), Persona Consistency (0.88 $\rightarrow$ 0.90), and Ethical Behavior (0.88 $\rightarrow$ 0.93), suggesting traffic steering provides clearer behavioral anchors than PRB allocation. Conversely, CodeAgent-RobustImplementer exhibits severe degradation (0.74 $\rightarrow$ 0.58, 22\% drop) driven by Linguistic Habits collapse (0.71 $\rightarrow$ 0.46, 35\% decline). RobustImplementer's defensive communication style—extensive error handling explanations and safety-focused terminology—conflicts increasingly with task complexity. Planner demonstrates divergent trajectories: Default improves (0.85 $\rightarrow$ 0.92, 8\% gain) with strong Persona Consistency enhancement (0.86 $\rightarrow$ 0.96), while CreativeThinker declines (0.67 $\rightarrow$ 0.59, 12\% drop) with severe Linguistic Habits and Persona Consistency erosion (both dropping to 0.46). The persistent 0.50 failure signature in Analyser-FastFailAuditor across both tasks indicates systematic persona rejection rather than task-specific breakdown. Prescriptive compliance emerges from complex interactions among persona characteristics, task demands, and architectural constraints. Certain combinations exhibit fundamental incompatibilities manifesting as detectable failure patterns, requiring pre-deployment identification through systematic validation.

\subsection{Behavioral Analysis}

\begin{figure}
    \centering
    \includegraphics[clip, trim=0cm 0cm 0cm 0cm,width=\columnwidth]{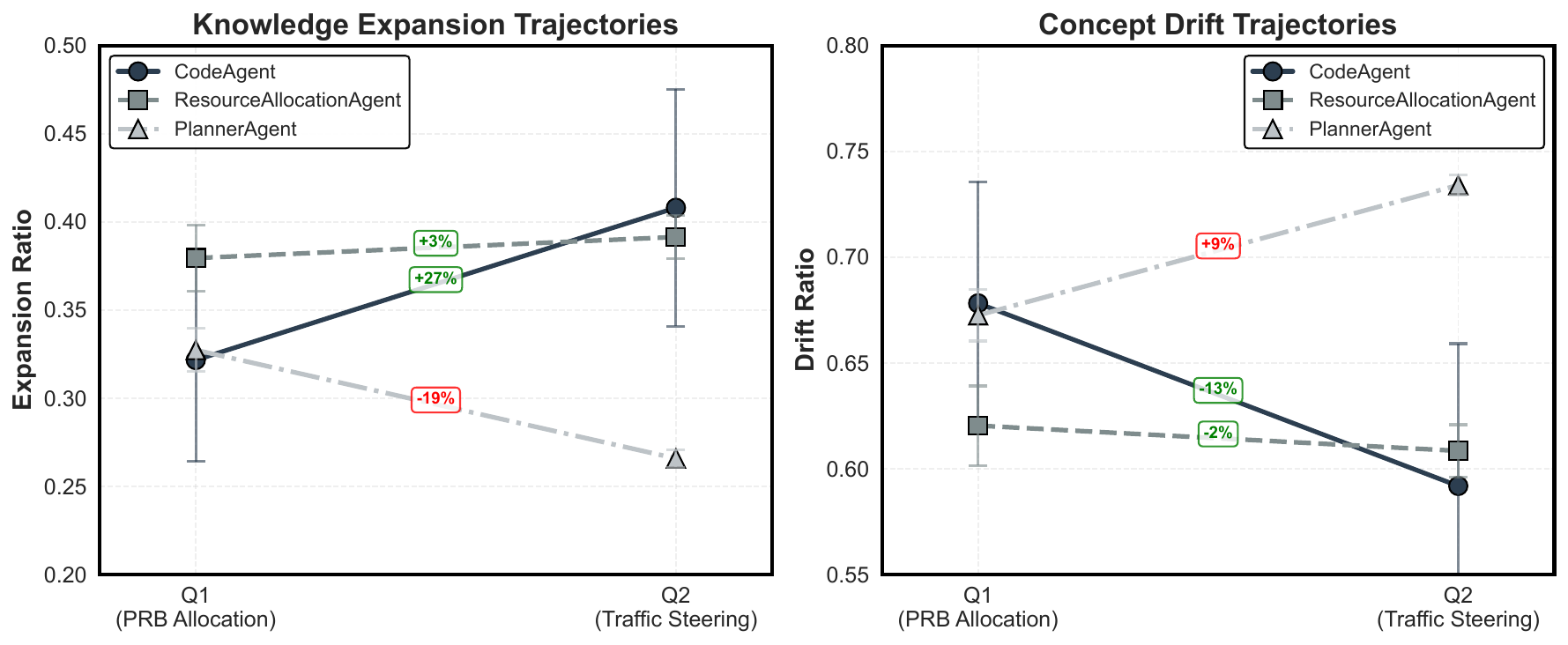}
    \caption{Knowledge expansion trajectories across Q1 and Q2. Higher expansion and lower drift indicate better ontology quality.}
    \label{fig:planercodeallocator}
\end{figure}

Fig.~\ref{fig:planercodeallocator} shows knowledge expansion across 486 persona configurations, measured by expansion ratio (relevant domain concept introduction) and drift ratio (off-topic concept introduction). Knowledge expansion patterns reveal striking task-dependent divergence. On Q1 (PRB allocation), ResourceAllocationAgent achieves highest expansion ($\mu = 0.380 \pm 0.019$), followed by PlannerAgent ($\mu = 0.327 \pm 0.012$) and CodeAgent ($\mu = 0.322 \pm 0.057$). These rankings shift substantially on Q2 (traffic steering): CodeAgent improves dramatically to $\mu = 0.408 \pm 0.067$ (27\% increase), ResourceAllocationAgent maintains stability at $\mu = 0.391 \pm 0.012$ (3\% increase), while PlannerAgent declines significantly to $\mu = 0.266 \pm 0.005$ (19\% decrease). These divergent trajectories demonstrate that task complexity affects knowledge expansion mechanisms differently across agents. CodeAgent's improvement reflects FlexRIC RAG retrieval stability—concrete implementation patterns provide consistent domain grounding regardless of algorithmic complexity. ResourceAllocationAgent's stability indicates GraphRAG retrieval maintains similar domain coverage across both resource optimization scenarios. In contrast, PlannerAgent's decline suggests multi-objective traffic steering reduces knowledge graph traversal relevance, with 73\% of introduced concepts falling outside expected telecom domains on Q2 versus 67\% on Q1. Notably, persona variations have minimal impact on expansion metrics—within each agent type, different personas produce nearly identical expansion and drift patterns, indicating that knowledge expansion is determined by agent role and retrieval architecture rather than behavioral prescriptions.

\begin{figure}
    \centering
    \includegraphics[clip, trim=0cm 0cm 0cm 0cm,width=\columnwidth]{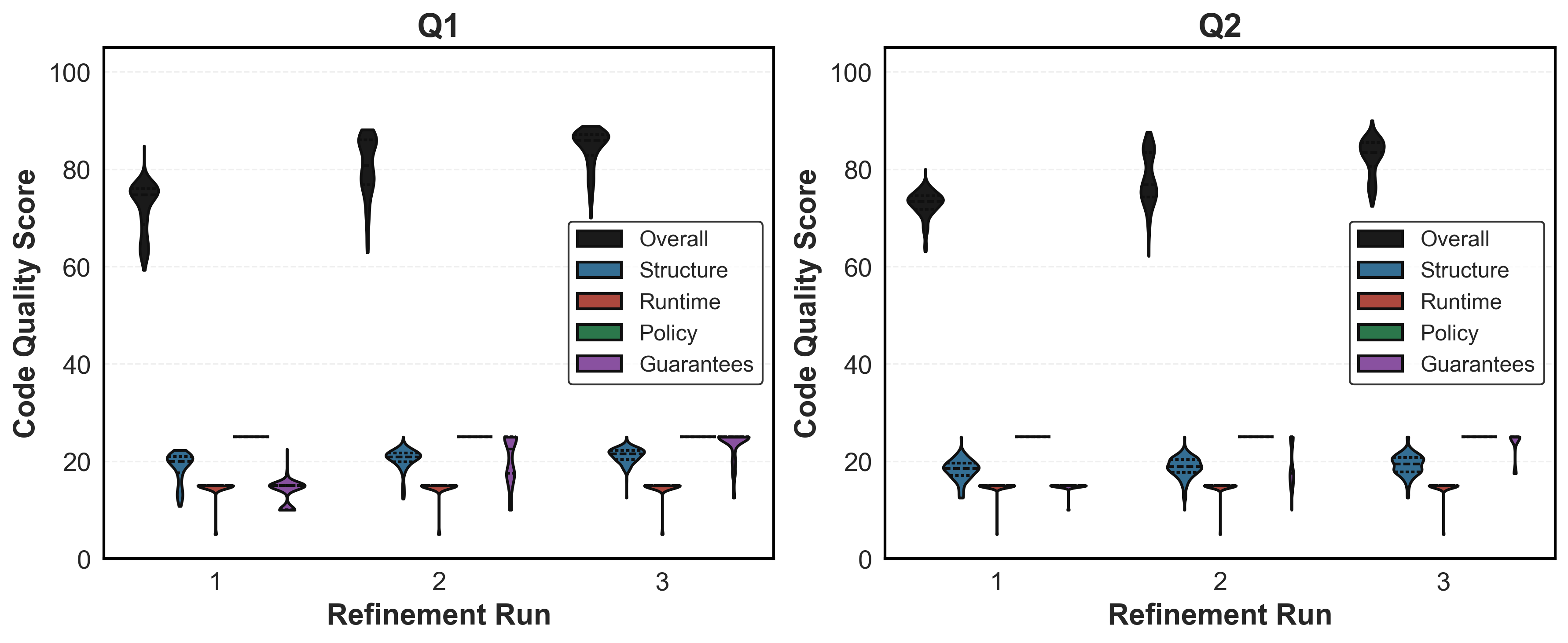}
    \caption{Code quality progression across three refinement runs. Overall score averages four equally-weighted components (25\% each): Code Structure, Runtime Behavior, Policy Compliance, and Formal Guarantees.}
    \label{fig:codequality}
\end{figure}

Fig.~\ref{fig:codequality} shows generated code quality across three refinement runs. Overall quality improves 15.1\%, from 72.5 in Run 1 to 83.4 in Run 3. However, components exhibit markedly different trajectories. Formal Guarantees show the most dramatic improvement, increasing from 56.9 to 94.0 (65.4\% improvement), indicating the refinement process effectively addresses formal verification requirements. Code Structure improves moderately from 74.1 to 81.0 (9.3\%). Policy Compliance remains constant at 100.0 across all runs, suggesting generated code consistently satisfies policy requirements from initial generation. In contrast, Runtime Behavior shows no improvement (58.9 $\rightarrow$ 58.8), remaining the primary bottleneck. These findings suggest that while multi-agent refinement effectively addresses structural and formal guarantee issues, runtime behavior optimization requires additional specialized attention or alternative refinement strategies.


\begin{figure}
    \centering
    \includegraphics[clip, trim=0cm 0cm 0cm 0cm,width=\columnwidth]{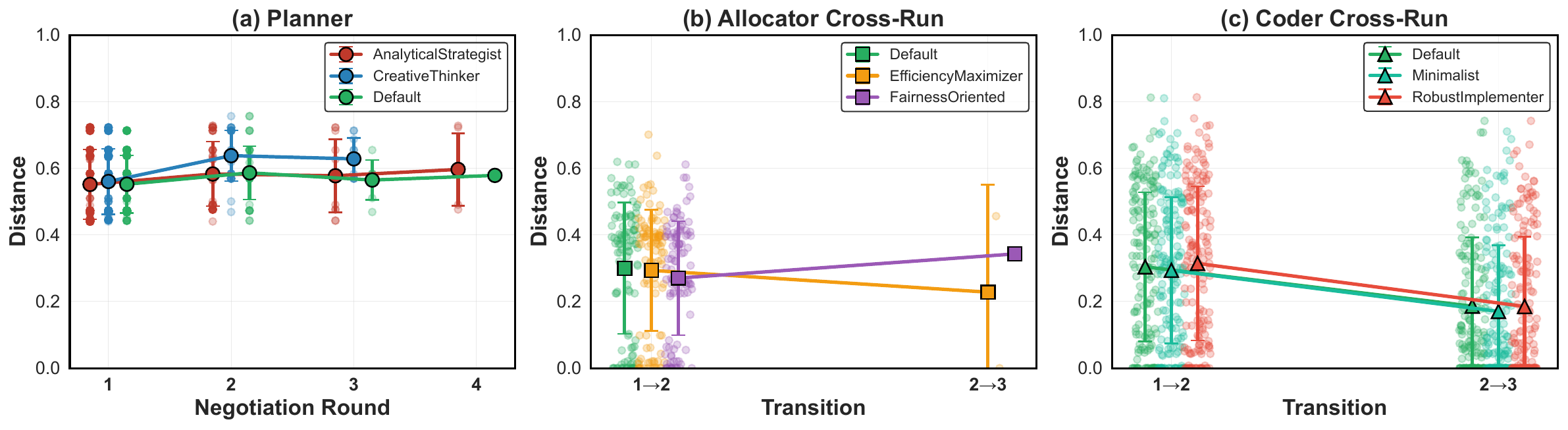}
    \caption{Semantic embedding distances: (a) Planner context divergence across rounds, (b) ResourceAllocator cross-run stability by persona, (c) Coder cross-run convergence by persona. Lower values indicate better alignment and stability.}
    \label{fig:combined_embedding_scatter}
\end{figure}

Semantic divergence in PlannerAgent reveals a fundamental tension in multi-turn negotiation systems. As negotiation rounds progress, distance between embedding context (retrieved domain knowledge) and problem-solution context (evolving negotiation state) increases 6.5\%, with task complexity amplifying this effect (Q2 shows 32.6\% worse alignment than Q1). This divergence suggests retrieval-augmented planning struggles to maintain semantic coherence when problem spaces expand through iterative refinement. Minimal persona impact (2.9\% gap between best and worst) indicates this challenge is architectural rather than behavioral—GraphRAG's graph traversal patterns inadequately track dynamic multi-agent negotiation evolution. Unlike static retrieval where context remains fixed, negotiation introduces temporal dependencies that current retrieval architectures fail to capture, causing PlannerAgent's embedding context to drift from the actual problem being solved.


In contrast, code generation and resource allocation demonstrate convergent refinement dynamics with markedly different trajectories. CodeAgent achieves 14\% greater semantic stability than ResourceAllocationAgent (0.242 versus 0.277) and exhibits dramatic two-phase behavior: initial exploration (Run 1 $\rightarrow$ 2: 0.304) followed by rapid convergence (Run 2 $\rightarrow$ 3: 0.181, $-$40\% acceleration). This pattern, consistent across all Coder personas (39--42\% acceleration), suggests xApp code generation benefits from FlexRIC's stable codebase structure—once initial implementation patterns are established, subsequent refinements converge toward verified templates. ResourceAllocationAgent's more modest convergence ($-$7\% Run 2 $\rightarrow$ 3) reflects continuous optimization in PRB allocation, where each refinement explores marginally different resource distributions rather than stabilizing toward fixed solutions. The 35\% perfect convergence rate for CodeAgent versus 0.3\% for ResourceAllocationAgent underscores this fundamental difference: code generation seeks deterministic correctness while resource optimization pursues dynamic efficiency.


Persona-driven performance patterns reveal that principle-based configurations achieve superior semantic consistency: Minimalist code generation (0.232) and FairnessOriented resource allocation (0.271) outperform Default implementations by 6--10\%. These improvements suggest explicit optimization principles—whether code minimalism or allocation fairness—provide semantic anchors that stabilize refinement trajectories. However, limited persona impact on PlannerAgent (2.9\% gap) versus stronger effects on CodeAgent (6\%) and ResourceAllocationAgent (10\%) indicates persona effectiveness depends on task structure. In well-defined generation tasks (code, resource plans), personas shape optimization strategies; in open-ended planning tasks, architectural limitations dominate. This distinction has critical implications for multi-agent system design: retrieval-augmented components require architectural innovation to handle temporal dynamics, while generation-focused components can leverage behavioral diversity through persona engineering to achieve specialized performance characteristics.


\subsection{Persona Impact Analysis}
\label{fig:persona_heatmaps}
\begin{table*}[t]
\centering
\caption{Persona impact averaged across Q1 and Q2 configurations. Deltas ($\Delta$) show percentage point changes from all-default baseline.}
\label{tab:persona_impact_avg}
\scriptsize  
\begin{tabular}{llcccc}
\toprule
\textbf{Agent} & \textbf{Persona} & \textbf{Normative $\Delta$ (\%)} & \textbf{prescriptive $\Delta$ (\%)} & \textbf{Behavioral $\Delta$ (\%)} & \textbf{Average $\Delta$ (\%)} \\
\midrule
\multirow{2}{*}{Planner} 
    & CreativeThinker & $-11.5$ & $-32.0$ & $+0.5$ & $-14.3$ \\
    & AnalyticalStrategist & $+0.2$ & $-1.3$ & $+1.3$ & $+0.1$ \\
\midrule
\multirow{2}{*}{Coordinator} 
    & Strategist & $+11.3$ & $+29.2$ & $+1.2$ & $+13.9$ \\
    & Tactician & $+20.0$ & $+0.8$ & $+1.5$ & $+7.5$ \\
\midrule
\multirow{2}{*}{Allocator} 
    & FairnessOriented & $+16.6$ & $-4.0$ & $-0.2$ & $+4.1$ \\
    & EfficiencyMaximizer & $+22.4$ & $+16.0$ & $-4.5$ & $+11.3$ \\
\midrule
\multirow{2}{*}{CodeAgent} 
    & Minimalist & $+10.2$ & $-10.2$ & $-0.2$ & $-0.1$ \\
    & RobustImplementer & $+1.7$ & $-14.5$ & $+1.5$ & $-3.8$ \\
\midrule
\multirow{2}{*}{Analyser} 
    & StrictAssessor & $-5.3$ & $-20.5$ & $-1.9$ & $-9.2$ \\
    & FastFailAuditor & $+11.4$ & $-29.2$ & $-0.6$ & $-6.1$ \\
\bottomrule
\end{tabular}
\end{table*}

Table~\ref{tab:persona_impact_avg} compares persona-customized agent configurations against an all-default baseline across three evaluation dimensions: normative compliance (rule adherence), prescriptive alignment (contextual understanding), and behavioral safety (code quality). The table shows \textbf{per-agent performance deltas} ($\Delta_{\text{agent}}$) in percentage points, where positive values indicate improvement and negative values indicate degradation. Each $\Delta_{\text{agent}}$ represents the change in a specific agent's performance when a persona configuration is applied, calculated as $\Delta_{\text{agent}} = \text{Score}_{\text{persona}} - \text{Score}_{\text{baseline}}$, capturing both direct effects (on the modified agent) and cascading effects (on other agents due to multi-agent dependencies). Additionally, \textbf{system-wide deltas} ($\Delta_{\text{system}}$) average performance changes across all five agents, providing a holistic measure of each configuration's impact on overall system behavior.


Persona customization yields highly heterogeneous impacts across agents and evaluation categories. The Coordinator demonstrates the strongest overall positive impact: \textit{Strategist} ($+11.3\%$ normative, $+29.2\%$ prescriptive, $+1.2\%$ behavioral) achieves $+13.9\%$ average improvement, while \textit{Tactician} ($+20.0\%$ normative, $+0.8\%$ prescriptive, $+1.5\%$ behavioral) achieves $+7.5\%$ average improvement. These results suggest multi-agent orchestration and task decomposition benefit significantly from personas emphasizing either strategic long-term planning or tactical short-term responsiveness. Conversely, Planner shows the most dramatic negative impact: \textit{CreativeThinker} ($-11.5\%$ normative, $-32.0\%$ prescriptive, $+0.5\%$ behavioral, $-14.3\%$ average) indicates overly creative exploration compromises adherence to formal reasoning structures in knowledge graph navigation.


Allocator demonstrates strong performance with both personas. \textit{EfficiencyMaximizer} achieves the highest normative ($+22.4\%$) and prescriptive ($+16.0\%$) gains among all configurations, though at the cost of behavioral degradation ($-4.5\%$), yielding $+11.3\%$ overall improvement. In contrast, \textit{FairnessOriented} shows more modest but balanced gains ($+16.6\%$ normative, $-4.0\%$ prescriptive, $-0.2\%$ behavioral, $+4.1\%$ average), suggesting resource allocation strategies can be optimized for specific objectives but involve trade-offs across evaluation dimensions. CodeAgent exhibits the most pronounced trade-off pattern: \textit{Minimalist} improves normative compliance ($+10.2\%$) while equally degrading prescriptive alignment ($-10.2\%$), resulting in near-zero net impact ($-0.1\%$ average). Similarly, \textit{RobustImplementer} shows substantial prescriptive degradation ($-14.5\%$) despite modest normative gains ($+1.7\%$), yielding $-3.8\%$ overall decline. Most critically, Analyser demonstrates consistent negative impacts across both personas: \textit{StrictAssessor} ($-5.3\%$ normative, $-20.5\%$ prescriptive, $-1.9\%$ behavioral, $-9.2\%$ average) and \textit{FastFailAuditor} ($+11.4\%$ normative, $-29.2\%$ prescriptive, $-0.6\%$ behavioral, $-6.1\%$ average) reveal that overly conservative or rapid-rejection analysis strategies severely constrain prescriptive understanding, suggesting excessive rigor in code evaluation inadvertently limits solution space exploration. These findings underscore that persona-agent alignment is not uniformly beneficial: while Coordinator personas yield improvements up to $+13.9\%$, misaligned personas in Planner and Analyser degrade performance by up to $-14.3\%$, highlighting the critical importance of matching cognitive styles to specific agent architectural roles.

Beyond individual agent performance, persona customization exhibits significant system-wide cascading effects (supplementary material). When a single agent's persona is modified, the impact propagates throughout the entire system, affecting all collaborating agents. Measured using $\Delta_{\text{system}}$, which averages performance changes across all five agents, configurations range from $+6.21\%$ (Coordinator: \textit{Strategist}) to $-0.74\%$ (Planner: \textit{CreativeThinker}), representing a 6.95 percentage point spread in holistic system behavior. Coordinator personas demonstrate the strongest positive cascading effects: \textit{Strategist} benefits 62.5\% of other agents ($+4.29\%$ average impact on non-Coordinator agents), while \textit{Tactician} benefits 87.5\% of other agents ($+3.94\%$ average impact), suggesting improved task orchestration and decomposition yield widespread downstream benefits across planning, resource allocation, code generation, and analysis phases. Conversely, Allocator: \textit{EfficiencyMaximizer}, despite achieving strong normative performance ($+22.4\%$) for its own agent, imposes net negative cascading effects on other agents ($-1.25\%$ average), degrading half of the system's components, indicating aggressive resource optimization can inadvertently constrain the solution space for dependent agents. These findings reveal that persona-agent alignment must be evaluated not only for local agent performance but also for \textbf{emergent multi-agent coordination dynamics}, as well-aligned personas in one architectural role can amplify system-wide performance through positive feedback loops, while misalignment can propagate degradation across the entire pipeline despite localized improvements.

\section{Conclusion}\label{sec:conclusion}

This paper introduces a persona-driven multi-agent framework for autonomous O-RAN network management with systematic evaluation enabling safe deployment in mission-critical telecommunications infrastructure. Analysis of 486 persona configurations across two O-RAN optimization challenges demonstrates that persona customization yields highly heterogeneous impacts—performance deltas ranging from $+13.9$\% (well-aligned) to $-14.3$\% (misaligned)—with single-agent modifications propagating system-wide through cascading coordination effects ($+6.21$\% to $-0.74$\% system-wide deltas).

Our decision-theoretic evaluation framework reveals four critical deployment insights: (1) retrieval architecture fundamentally constrains persona effectiveness—GraphRAG agents experience 19\% normative collapse on multi-objective tasks despite 27\% knowledge expansion, while RAG agents maintain stable criterion satisfaction with 40\% convergence acceleration; (2) optimal alignment is role-specific, with Coordinator personas achieving $+13.9$\% improvements while misaligned Planner personas degrade by $-14.3$\%; (3) aggressive local optimization degrades 50\% of system components through cascading effects despite individual gains; (4) systematic evaluation detects fundamental incompatibilities (0.50/0.50/0.50 failure signatures) before deployment.

Future work should explore predictive models for cascading effects, expand persona libraries with domain-specific archetypes, and investigate dynamic persona adaptation.  Our framework generalizes beyond O-RAN to any multi-agent domain requiring formal constraints, contextual understanding, and safety guarantees—including autonomous vehicles, distributed resource management, and collaborative robotics. As LLM-based systems become prevalent in critical infrastructure, this work provides the validation mechanisms necessary for reliable, safe deployment in mission-critical environments.

\section*{Acknowledgment}
This work was supported by UKRI851 on AI-agent cooperation for telecom safety and governance and EPSRC grants EP/Y037421/1 (CHEDDAR) .

\bibliographystyle{unsrt}
\bibliography{bib_template}

\end{document}